# Bifurcating-Paths: the relation between preferential flow bifurcations, void, and tortuosity on the Darcy scale.

Avioz Dagan[1], Yaniv Edery[1]

[1] Faculty of Civil and Environmental Engineering, Technion, Haifa, Israel

## Abstract

In recent years, Darcy scale transport in porous media was characterized to be Fickian or non-Fickian due to the homogeneity or heterogeneity of the porous medium conductivity layout. Yet, evidence shows that preferential flows that funnel the transport occur in heterogeneous and homogenous cases. We model the Darcy scale transport using a 2D conductivity field ranging from homogenous to heterogeneous and find that these preferential flows bifurcate, leaving voids where particles do not invade while forming a tortuous path. The fraction of bifurcations decreases downflow and reaches an asymptotical value, which scales as a power-law with the heterogeneity level. We show that the same power-law scaling of the bifurcations to heterogeneity level appears for the void fraction, tortuosity, and fractal dimension analysis with the same heterogeneity level. We conclude that the scaling with the heterogeneity is the dominant feature in the preferential flow geometry, which will lead to variations in weighting times for the transport and eventually to anomalous transport.

## 1. Introduction

Transport in saturated porous medium has been extensively investigated over the last few decades, with implications in many disciplines like chemical transport in soils and fractured rocks [*Bear*, 2013; *Haggerty et al.*, 2001; *Raveh-Rubin et al.*, 2015], oil recovery [*Edery et al.*, 2018; *Lenormand et al.*, 1983], filtration [*Tufenkji and Elimelech*, 2004], fuel cells [*Pharoah et al.*, 2006], and even percolation of coffee [*Fasano and Talamucci*, 2000]. The general advective-diffusion equation describes Fickian transport in a homogenous porous material, using:

1. $$\frac{\partial c}{\partial t} - \nabla \cdot (D\nabla c) + \nabla \cdot (vc) = S,$$

where c is the tracer concentration, D is the diffusion constant, v is solvent velocity, and S is the dissolved source that includes the concentration injected into the field [*Koch and Brady*, 1987]. However, a natural porous environment is characterized by a heterogeneous structure of the medium, which leads to non-Fickian transport with a heavy tail for the tracer migration that can be captured by various models [*Berkowitz and Scher*, 1997; *Cirpka and Kitanidis*, 2000; *Cushman and Ginn*, 1993; *Dullien*, 2012; *Haggerty et al.*, 2000; *Kang et al.*, 2011; *Le Borgne et al.*, 2008; *Morales-Casique et al.*, 2006a; b; *Sánchez-Vila and Carrera*, 2004; *Willmann et al.*, 2008]. This transport in a heterogeneous porous media can be regarded as transport within locally homogenous areas, distributed in a lognormal way, and spatially varying following a correlation length [*Gómez-Hernández and Journel*, 1993], while the variance of the distribution marks the field heterogeneity

[*Sanchez-Vila et al.*, 2006]. This heterogeneity formation allows solving the transport as Fickian on the local scale while giving rise to a non-Fickian transport in the field scale [*Edery*, 2021; *Edery et al.*, 2014]. A persistence outcome of the heterogeneity, both experimentally and numerically, is the emergent preferential flow paths defined as the tracer's movement in unequal parts through the porous medium [*Cirpka and Kitanidis*, 2000; *Willmann et al.*, 2008]. In these preferential flow paths, the fluid funnels into narrower flow paths with the lowest resistance to flow, and increasingly forms stagnation areas, or voids, in regions with high resistance [*Webb and Anderson*, 1996]. This relation between preferential flow and hydraulic conductivity distribution is observed at the field scale [*Bianchi et al.*, 2011; *Edery et al.*, 2016a; *Riva et al.*, 2010; *Riva et al.*, 2008], numerically [*Fiori and Jankovic*, 2012], and even at the pore-scale. Moreover, studies have shown the importance of preferential flows to the reaction pattern in reactive transport [*Edery et al.*, 2011; *Edery et al.*, 2015; *Edery et al.*, 2016b; *Raveh-Rubin et al.*, 2015]; and specifically for the vadose zone they are related to contaminant distribution in soil [*Hagedorn and Bundt*, 2002], microbial communities in soil [*Bundt et al.*, 2001; *Morales et al.*, 2010], and even landslides [*Hencher*, 2010; *Shao et al.*, 2015]. Preferential flows pattern starts as a uniform tracer front, which funnels and splits into distinct flowing branches, sometimes referred to as channel branching [*Fiori and Jankovic*, 2012; *Liao and Scheidegger*, 1969; *Moreno and Tsang*, 1994; *Torelli and Scheidegger*, 1972]. As the channel branches, the tracer concentrations vary and sample the conductivities non-uniformly. This branching is similar to single and multiphase flow at the pore scale [*Ferrari et al.*, 2015; *Yeates et al.*, 2020], and with a topology analogous to graph theory [*Kanavas et al.*, 2021; *Liao and Scheidegger*, 1969; *Torelli and Scheidegger*, 1972].

We identify the channels as the continuous transport of at least one particle in our particle tracking (PT) model, while the point at which a channel branch is a bifurcation of the transport. This bifurcation phenomenon was studied in similar fields, such as small-scale heat transfer bifurcation in porous media [*Yang and Vafai*, 2011a; b], and bifurcations in braided rivers [*Amooie et al.*, 2017; *Bolla Pittaluga et al.*, 2003; *Zolezzi et al.*, 2006]. However, to date, there is no study characterizing the bifurcation of flow in porous media at the Darcy scale and in the context of preferential flow paths. This work characterizes the preferential flow patterns and bifurcation on this Darcy scale transport. We identify the bifurcation points and the stagnant zones (voids) in a numerical conductivity field ranging from homogenous to heterogeneous. We further show that the bifurcations decrease from inlet to outlet, reaching an asymptotical value that scales with the heterogeneity level. We identify a power-law scaling that correlates the bifurcation and void fractions with the heterogeneity level. Surprisingly, the same power-law scaling with heterogeneity holds for the transport tortuosity and characteristic fractal dimension.

## 2. Methodology

We characterize the bifurcation of preferential flows using a set of 2D numerical simulations, where a second-order stationary random field of conductivities is distributed by a lognormal distribution with mean $\ln(k) \sim 0$ and a variance of $\sigma^2 = 1 - 5$, established by a sequential Gaussian simulator (GCOSIM3D) [*Gómez-Hernández and Journel*, 1993]. This conductivity field is made of 120×300 conductivity bins (each with a size of $\Delta = 0.2$ cm). Each field is produced by a statistical homogenous and isotropic Gaussian field in the $\ln(k)$, with a normalized correlation length $l/L = 0.016$, where $L$ is the domain length along the main flow direction and $l=1$ is produced by an exponential covariance. This correlation length leads to $\Delta/l = 0.2$, which provides an accurate description for the small-scale fluctuations generated by the ln(k) field and advective transport [*Ababou et al.*, 1989; *Riva et al.*, 2009]. One hundred realizations are produced for each variance with a stability test to verify the mean conductivity for all the results presented here; see supplementary for details. Each realization had a deterministic pressure drop translated to the total head drop ($H_{inlet-outlet} = 100$ cm) imposed from the inlet (left) to the outlet (right), and a finite element numerical model with Galerkin weighting function calculated the local head drop in 2D per bin [*Guadagnini and Neuman*, 1999]. Thus, the streamline for each realization is retrieved, and from it, the local velocities, given that the porosity θ=0.3. Each PT realization tracks a pulse of $10^5$ particles. These particles are flux-weighted at the inlet according to the inlet permeability distribution, and at t=0, the particle pulse advances according to the local advection and diffusion term, following the Langevin equation:

2. $$d = v[x(t_k)]\delta t + \boldsymbol{d}_D,$$

where d is the displacement, $x(t_k)$ is the particle known location at time $t_k$, $v$ is the fluid velocity at that location, $\delta t = \frac{\delta s}{v}$ is the temporal displacement magnitude ($v$ is the modulus of v) and $\boldsymbol{d}_D$ is the diffusive displacement. The displacement size δs is selected to be an order of magnitude less than Δ so to interpolate the velocity within each bin correctly. This diffusive displacement is randomly generated from a normal distribution between 0 to 1 (N[0,1]), multiplied by the square root of the diffusion coefficient ($D_m = 10^{-5} \frac{cm^2}{sec}$) representing the diffusion of ions in water [*Domenico and Schwartz*] and the displacement magnitude as illustrated from the following equation:

3. $$d_D = N[0,1]\sqrt{2D_m \delta t}.$$

The simulations were verified using $10^6$ particles and δs <Δ/10 with no significant numerical dispersion. Using the same GCOSIM3D model was proved valuable in reproducing and analyzing field data [*Eze et al.*, 2019; *Obi et al.*, 2020], uncertainty[*Ciriello et al.*, 2013; *Franssen et al.*, 2004; *Riva et al.*, 2005], and upscaling [*Li et al.*, 2011], while the PT method proved to be very robust and appropriate in this modeling configuration [*Peter Salamon et al.*, 2006; *P Salamon et al.*, 2006].

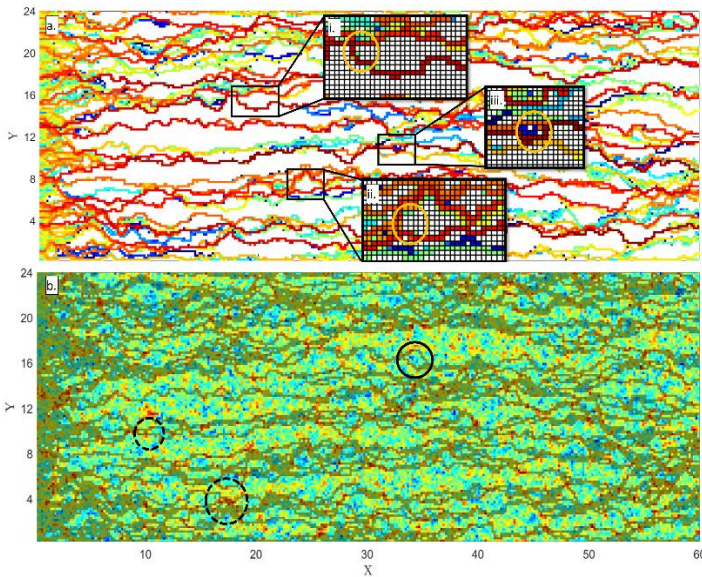

Figure 1. a. Example of particles visitations per bin, where a heat map represents the particle visitations throughout the simulation, on a logarithmic scale, white areas mark locations with no particle's visitation. i. Example of a single bifurcation point where there is no visitation of particles in the bifurcation point and downstream from it, and there are visitations above and below the bifurcation points. ii. An example of a bifurcation front, where multiple cells have no visitations. iii. An example for a single stagnant point which is not defined as a bifurcation point. b. Combined description of the particle visitation overlay on the conductivity field of a single realization with variance of ln(k)= 5, and a color bar for ln(k). The darker hue describes the flow paths, while the complete circle marks a bifurcation around a low conductivity, the small, dashed circles marks a higher-than-average conductivity in which there is no transport, and the bigger dashed circle marks a higher-than-average conductivity that funnels the transport.

While defining the bifurcation as the point where a channel branches into two is straightforward, searching for the bifurcations within the preferential flows is more challenging. We, therefore, distinguish a set of points where the channel first splits while changing the flow direction tangentially. Defining the bifurcation point as a cell with transport up-flow from it and no transport at down-flow from it, yet there is transport transversely to the down-flow direction (see example in figure 1. a.i). However, in some cases, the bifurcation does not occur at a single cell but rather at several vertically consecutive cells (= 'bifurcation front') as shown in figure 1.a.ii. The bifurcation front is typically around 1% of the total bifurcations and a maximum of 8% of the bifurcations per realization. This work does not consider a bifurcation front since they are negligible and uncertain in conductivity value. In addition, as shown in figure 1.a.iii, individual voids points are not defined in this work as bifurcation points, although locally, these are points that split the flow.

3. **Results**

In the following, we identify and characterize the bifurcation from inlet to outlet for fields with varying heterogeneity levels. While many parameters may affect transport on the Darcy scale, e.g., isotropy, correlation length[*Edery*, 2021], and heterogeneity [*Edery et al.*, 2014], we focus on the transition from a homogenous to a heterogenous case; and show how this transition affects the topology of preferential flows. We demonstrate that a power-law correlates the bifurcation to heterogeneity. The same power-law correlates the void, the tortuosity, and the fractal dimension of the preferential flows to the heterogeneity of the field.

While the definition of bifurcations is established for this study, the mechanism leading to these bifurcations is not. In the simplified case where there is a non-conductive area or a big difference between conductivities, it is evident that a bifurcation will occur around the obstacle, which makes it a local event. However, bifurcations mainly occur upstream to the location with the low conductivity in a way that maximizes the path of least resistance globally. Evidence for this kind of bifurcation around a low conductive area can be found in Levy and Berkowitz [*Levy and Berkowitz*, 2003], where the tracer is branching around low permeable zones, yet some tracer does enter these low conductive zone, as it serves the overall minimization of flow through the global solution of transport as shown in the recent study by [*Zehe et al.*, 2021]. This branching of transport is presented in the supplementary material (figure S1 modified from Levy and Berkowitz, 2003; © with permission from Elsevier 2003), and the temporal and spatial evolution of the pulse tracer exhibits similar behavior as in our simulations, as shown by [*Berkowitz*, 2021]. The preferential flows are marked by a substantial spatial change of the tracer concentration from inlet to outlet. We encounter these preferential flows in all scales, from the pore scale to the field. Yet, the flow resistance mechanism that forms them changes from variations in the surface area to volume at the pore scale, to variations in conductivity at the Darcy\field scale.

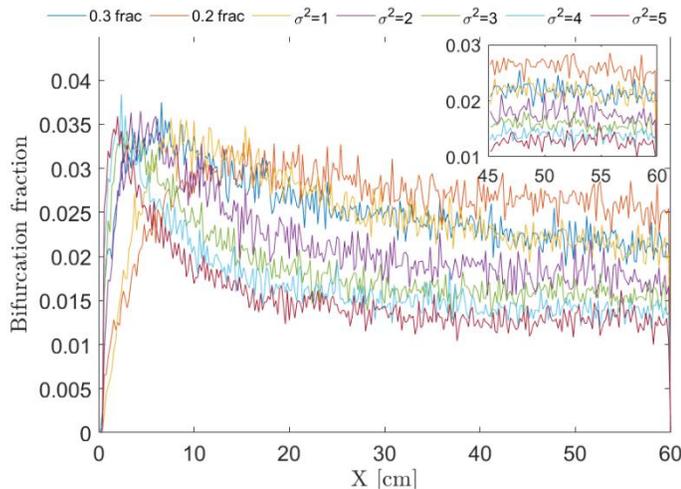

*Figure 2. Bifurcation fraction from all the cells along the y-axis, and their change downflow on the x-axis. Each color represents a different heterogeneity level, yet all reach an asymptotical value, shown in the inset.*

These preferential flows maximize their presence at higher conductivities while bifurcating around low conductivities (figure 1. b), forming a path of least resistance. As such, the ratio between the conductivity at the bifurcation point and the point downflow where there are no particles should be consistently higher than one. Yet, the histogram of bifurcation ratio has a substantial portion smaller than one (figure S2 in the supplementary), pointing to a global and not local minimization of energy of the transport. This histogram ratio, which is on average smaller than one, scales with the variance of the domain (figure S3 in the supplementary) for all the ensemble realizations. An additional mechanism that may modify the bifurcation ratio is the Peclet number ($Pe$), defined as:

4. $$Pe = \frac{V_d \sqrt{k}}{n_e D_d} = \frac{V_d \sqrt{K \frac{\mu}{\rho g}}}{n_e D_d}$$

where $V_d$ is Darcy velocity; $k$ is the intrinsic permeability; $n_e$ is effective porosity; $K$ is (arithmetic) average; $\mu$ and $\rho$ are dynamic viscosity and density of water, respectively; $g$ is gravity; and $D_d$ is diffusion. The ensuing value of $Pe$ is indeed indicative of an advection-dominated transport setup (see [*Huysmans and Dassargues*, 2005] for details). Changing the head drop ($H_{inlet-outlet}$) from 100 to 10 and 1cm, results in changing the Darcy velocity ($V_d$) from 5.5 to 0.5 and 0.05, and the $Pe$ from 597 to 59.7 and 5.97 respectively. When examining the conductivity upflow to the bifurcation (blue dashed line) and the conductivity downflow from the bifurcation (black solid line) compared to the field conductivity distribution (red circles), the results are only slightly different as the $Pe$ reduces (see figure S2 in the supplementary). However, we anticipate that in a extremely diffusion-dominated system, preferential flows, and therefore bifurcations, will become less dominate, until disappearing due to the growing diffusion component.

A closer look at figure 1. b., and the histograms in figure S2 in the supplementary, shows that there are areas where the transport occurs within the low conductivity zones, and therefore bifurcations do not generally follow a conductivity ratio higher than one. Furthermore, there are more bifurcations up-flow than downflow, following the continuous funneling that the preferential flows undergo, making them more distinct downflow. This issue is also addressed in [*Zehe et al.*, 2021], showing the declining entropy towards an asymptote due to the further funneling of preferential flows, a process acerbated with heterogeneity increase. This downflow reduction in the mean bifurcation is apparent in the bifurcation fraction over the Y-axis, averaged over the ensemble, with an initial increase in the number of bifurcations, followed by a decrease towards an asymptote for all variances (see figure 2 for details).

This bifurcation fraction from inlet to outlet also occur for a binary case, where we maintain an average conductivity for all the simulated fields and distribute a conductivity 6 orders of magnitude lower over 20% and 30% of the field. The bifurcation fraction for the conductivity field with a variance of 1 follows the binary distribution of 30%, while the 20% has a higher asymptote, with a bifurcation fraction closer to the ratio of the low conductive cells in the field, thus making it closer to a local event. An analysis of the binary case solely is presented in the supplementary, showing that this bifurcation scales with the increasing fraction of impervious area. Moreover, the maximum bifurcation fraction is 0.035 for all variances, yet, the asymptotical value becomes smaller as the variance increase, indicating that the funneling toward distinct preferential flows is stronger while the number of preferential flows decreases (see inset in figure 2 for details).

The bifurcation fraction asymptotical value decreases with the increase in variance, following a power law of: $f_{bif} = 0.026 \cdot (\sigma^2)^{-0.25} - 0.004$, where $f_{bif}$ is the fitted bifurcation fraction (see figure 3. a). While the prefactor (0.026) for the power law represents the maximum asymptotical value for the bifurcation ratio over the y-axis, the constant (0.004) is tied to the diffusion coefficient which allows particles to migrate outside the preferential flows. This scaling of the bifurcations to an asymptotical value occurs within ten realizations, and does not vary much from there on, as can be seen in figure S5 in the supplementary, while the variance among the bifurcation fraction decreases with the bifurcation fraction variance. This reduction in variance follows the streamline analysis in figure S4 in the supplementary that shows how the number of preferential flows and streamlines number decreases with the log-normal conductivity field variance.

By the definition of the bifurcation, each bifurcation leads to the formation of a void where there is no particle visitation or an extremely low tracer concentration, as can be seen in figure 1 and in experiments (see figure S1 in the supplementary) [*Berkowitz*, 2021; *Levy and Berkowitz*, 2003]. As such, there should be a relation between the void fraction in the domain and the variance. Indeed, the void fraction increases with the variance mirroring the bifurcation fraction trend (figure 3. a). This increase in void fraction follows the same power as that of the bifurcation fraction pattern, namely: $f_{Void} = -0.57 \cdot (\sigma^2)^{-0.25} + 1$, where $f_{Void}$ is the fitted void fraction. Here the asymptotical value (1) is for infinite heterogeneity that forms a single channel from inlet to outlet with no bifurcation and maximum void, while the prefactor (0.57) moderate this approach. This analysis points to the relation between the bifurcation and the formation of voids, or areas where transport is diffusion dominated, which scales with the variance of the domain. The fact that there is a void fraction in a Darcy scale simulation that scales with the variance points to the fact that the preferential flows are tortuous, even on the Darcy scale.

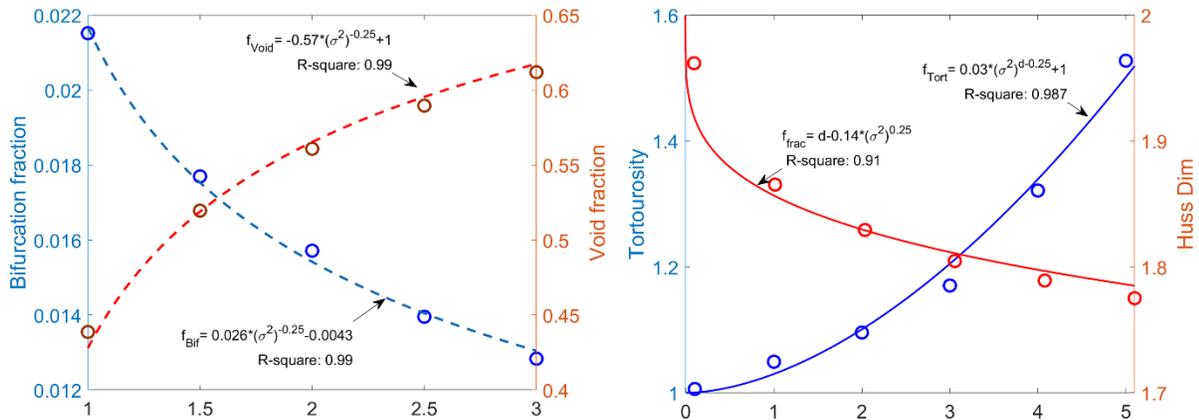

*Figure 3.a. Mean bifurcation fraction, taken from the asymptotical value in figure 3, as conductivity variance increases (blue circles) and a power-law fit (blue dashed line). Mean void fraction (red circles) as conductivity variances increases, with its power-law fit (red dashed line). b. Mean tortuosity value, calculated from particle trajectory, as conductivity variances increases (blue circles), and its power-law fit (blue dashed line). Mean Haussdorf fractal dimension as conductivity variances increases (red circles), and its power-law fit (red dashed line).*

The concept of tortuosity has been developed for the pore-scale to link porosity and pore structure to permeability [*Carman*, 1939; *Kozeny*, 1932]. Since then, numerous studies have tried to relate tortuosity to porosity and pore geometry, as summarized in the following review paper [*Ghanbarian et al.*, 2013], yet to the best of our knowledge, this concept was never generalized to a Darcy scale model or related to the heterogeneity of the medium. We calculate the tortuosity by measuring the trajectory length for each simulated particle and dividing it by the field's length ($D_i/L$). We do so for all particles and get the mean tortuosity for each realization ($f_{Tort} = (\sum_{i=1}^{n} D_i/L)/n$). This Darcy scale tortuosity is the outcome of the void and bifurcation, and therefore it follows the same scaling: $f_{Tort} = 0.03 \cdot (\sigma^2)^{d-0.25} + 1$, where 1 is the tortuosity for a completely uniform field ($\sigma^2 = 0$), and $d = 2$ is the dimension of the field, while the prefactor moderates the asymptote (see figure 3. b.).

The relation between tortuosity and heterogeneity suggests that the least resistance path scales with this power law's heterogeneity. This scaling is different from the tortuosity scaling for the pore-scale since it relates to the heterogeneity level and not the porosity, yet studies often use a fractal dimension to connect tortuosity to permeability [*Xia et al.*, 2018; *Yu and Cheng*, 2002; *Yu and Liu*, 2004]. Following this pore-scale concept, we calculate the Hausdorff fractal dimension (Hfd) using the box count method. To calculate the Hfd, images, like figure 1. a, were switched to a binary image, where we marked each bin with a particle visitation bigger than one as 1, or 0 if no particle visited. This binary image is divided into a number of boxes of equal size, and the sum of boxes with at least one 1 is extracted. This is repeated for all possible box sizes, which form a grid of evenly distributed boxes, and the slope relating the logarithm for the sum of 1's divided by the logarithm of the total boxes is the Hfd. This calculation is done for all the ensembles per variance, providing a mean Hfd for each variance. As can be seen in figure 3. b, the relation between the Hfd to the variance follows the same scaling: $f_{Huss} = d - 0.14 \cdot (\sigma^2)^{0.25}$, where $d = 2$ is the dimension of the problem. The variance for the bifurcation fraction, void fraction, tortuosity and Hfd is very narrow and convergence to its value very fast with each added iteration for the simulation ensemble, as shown in figure S5 in the supplementary.

4. **Summary and Discussion**

In this study, we define and analyze the role of transport bifurcation and show that while transport and bifurcation follow higher-than-average conductivities, bifurcations do also occur in high conductive areas, following the path of least resistance for the total transport and not the local scale. However, we show that the bifurcations fraction scales with heterogeneity through a power law, and as each bifurcation leads to the void formation, the same exponent also represents the relation between the void fraction and the heterogeneity. While these parameters are extracted for the field scale, the tortuosity is calculated directly from each particle route, yet it independently follows the same scaling exponent found in the previous analysis. We follow this analysis with a Hfd fractal dimension analysis on the transport and show that the same exponent scaling holds. A similar analysis was performed for the binary case, with impervious fraction, offering the same exponent and scaling holds for this case, pointing to the robustness of this scaling.

The exponent value for the power law which defines this emerging scaling, is the only value that fits the relation between the heterogeneity to the bifurcation, void fraction, tortuosity, and Hfd, leaving but one fitting parameter for each preferential flow characteristic. Yet the fact that there is an exponent common to all preferential flow's characteristics indicate that the optimization of transport in a conductive field leads to a geometrical pattern of preferential flow manifested in these parameters. This study aims at defining these preferential flows characteristics, and while the exact relations between these parameters are unknown, the framework to analyze them is presented. As numerical [*Edery et al.*, 2014] and experimental [*Levy and Berkowitz*, 2003] studies on transport are dedicated to relating the heterogeneity to the Continuous Time Random Walk (CTRW) framework, specifically to the $\beta$ exponent in the truncated power law (TPL) probability density function which defines the transition times in the CTRW-TPL, we will relate the TPL-CTRW parameters to the power-law scaling found here in a subsequent study.

We do expect that increasing the dimension to a 3D model will provide similar results that follow a power-law scaling, yet not with the same exponent. Similar to Percolation theory, when the bond percolation on a square lattice decreases from 0.5 in 2D to 0.248 in 3D (as proven by [*Kesten*, 1982]), or from a site percolation of 0.59 in 2D to 0.311 in 3D. While this change is dependent on the lattice used, it is persistent in nature as we expect from our system. Although our simulations do not deal with percolation, there are similarities that many researchers pointed out [*Hunt and Sahimi*, 2017], while experimental indications to the increased dominance of preferential flows is known [*Bianchi et al.*, 2011]. Therefore, we expect the bifurcation fraction versus X-axis compered to void fraction versus X-axis, to remain the same in nature, only having reduced value for the bifurcations and an increased value for the void since the possibility for a preferential flow to bypass low conductivities and optimize the path of least resistance has increased as we move from a 2D to a 3D system, similar to the percolation threshold.

While the tortuosity, void, and bifurcations are apparent in experimental setups, as shown in [*Levy and Berkowitz*, 2003] (figure S1) and [*Bianchi et al.*, 2011], the preferential flow pattern is not characterized experimentally in the literature since it is challenging to quantify a tracer pulse trajectory in a Darcy scale experimental setup. We propose a set of experiments, similar to [*Edery et al.*, 2015], where the tracer concentration is tracked in time on the Darcy scale, but the continuous point source of the tracer is replaced with a tracer front pulse. And to complicate things further, the domain must be made of distinguishable bead packings patches with known, homogenous conductivity, yet these patches' conductivity must vary spatially, mimicking the variance change presented here. While this experimental setup is feasible, it is extremely consuming and therefore not in the scope of this study. Nonetheless, the importance of these findings is in characterizing the topology of preferential flows through the concept of bifurcations, voids, and tortuosity at the Darcy scale and showing that these concepts scale with heterogeneity. This scaling may have implications for many disciplines of porous media research, from contaminant transport [*Edery*, 2021; *Edery et al.*, 2016a; *Edery et al.*, 2014; *Hagedorn and Bundt*, 2002; *Zehe et al.*, 2021], where the contamination location is dependent in the preferential flow pattern, to reactive transport, where the reaction is set by the reactant concentration that follows the preferential flows [*Edery et al.*, 2011; *Edery et al.*, 2015; *Edery et al.*, 2021; *Edery et al.*, 2016b; *Raveh-Rubin et al.*, 2015]. Moreover, there are indication that bacteria in soils are not only relying on these preferential flows [*Bundt et al.*, 2001], but actually forming them [*Morales et al.*, 2010].

# Bifurcating-Paths: the relation between preferential flow bifurcations, void, and tortuosity on the Darcy scale.

supplementary material

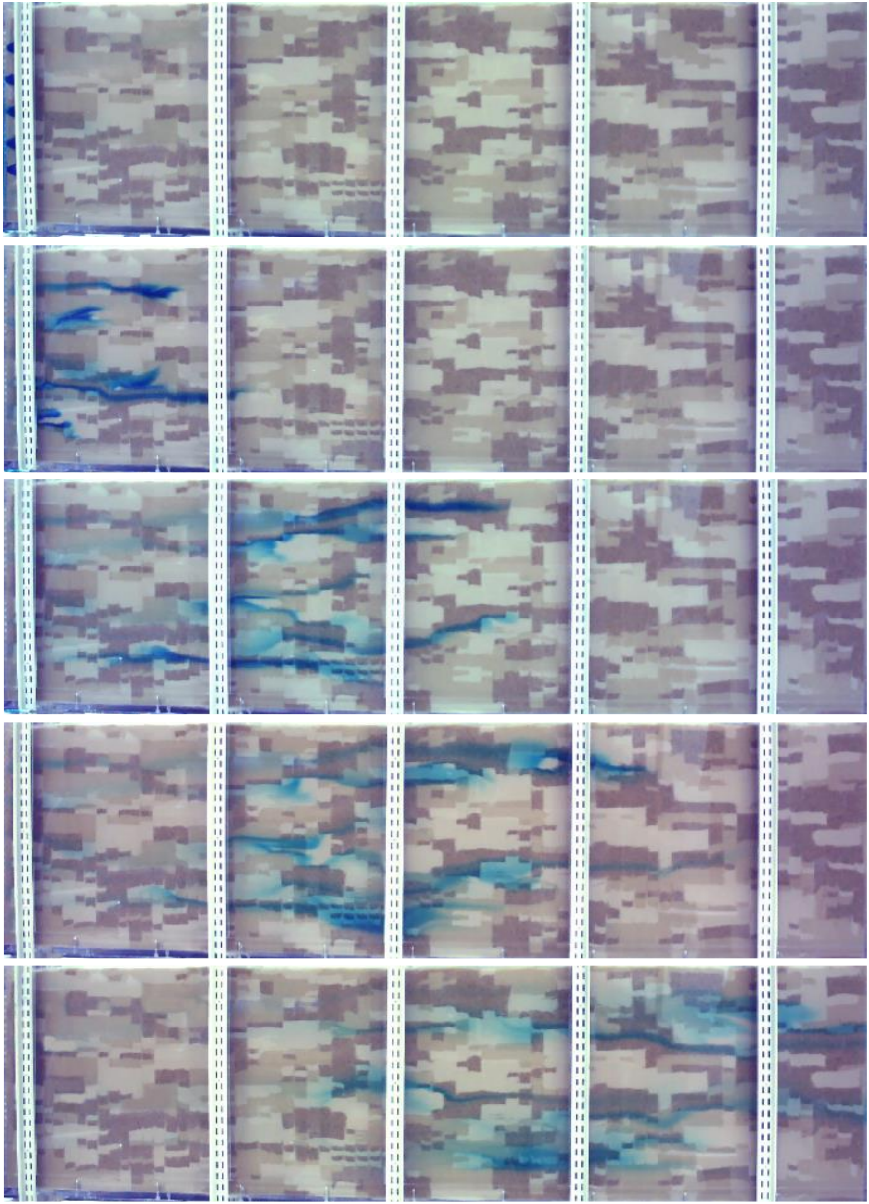

*Figure S1. Photographs of the randomly heterogeneous medium with five dye tracer point injections being transported, under constant flow from left to right, as seen through the front wall. The brown, light brown and white depict high, medium and low conductivity sands distributed in an exponentially correlated structure, in a flow cell of 213×65×10 cm.*

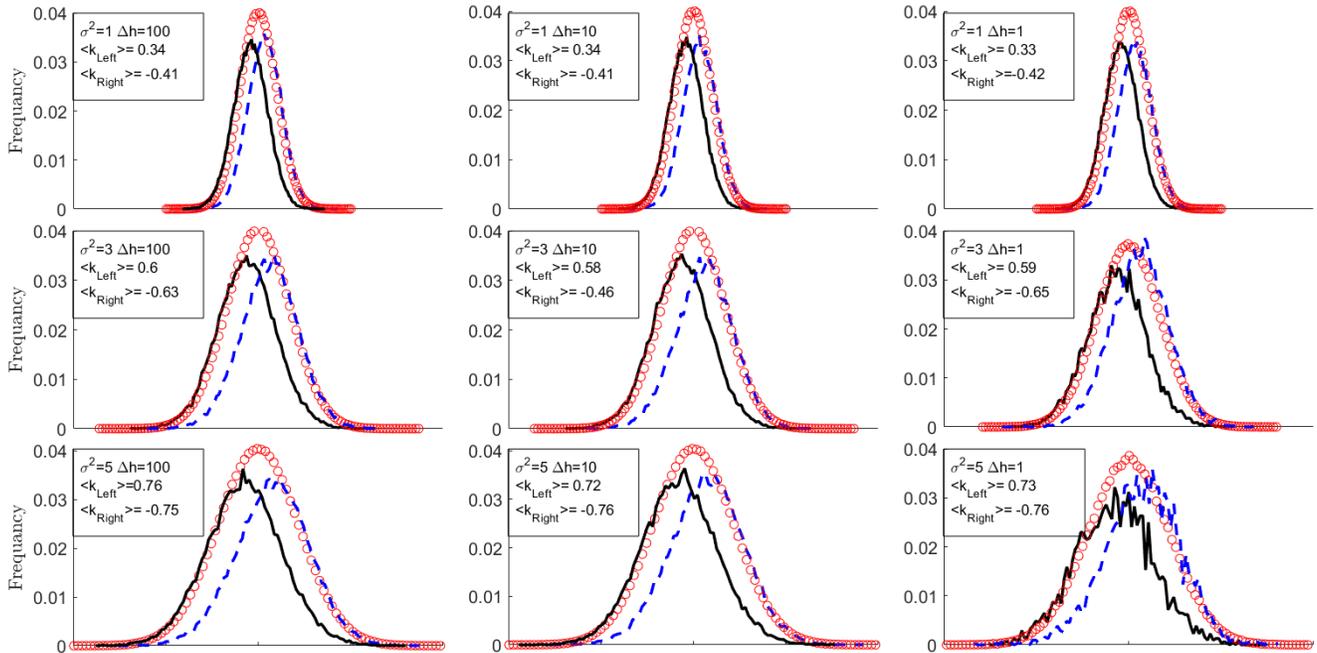

*Figure S2. Frequency for the log-normal conductivity for fields ensamble (red circles), the conductivity upflow to the bifurcation (blue dashed line) and the conductivity downflow from the bifurcation (black solid line), see box for the average values. Results correspond to $\Delta h =$ 100,10,1 (left, middle and right columns, respectively), and to $\sigma_0^2 = 1,3,5$ (upper, middle and bottom rows, respectively).*

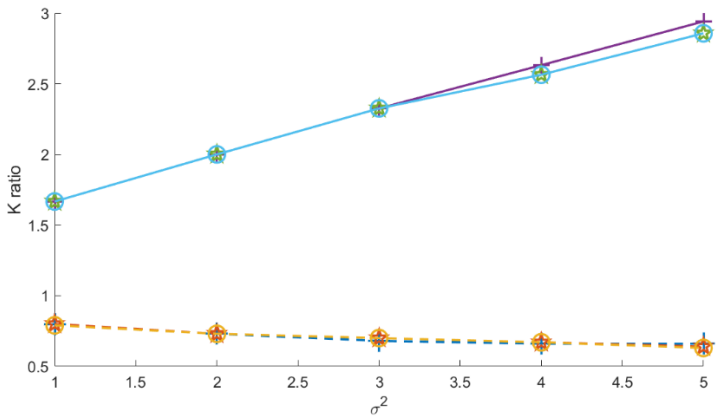

*Figure S3. The hetrogeneity magnitude ($\sigma_0^2$) effect on the mean of conductivity ratio between the bin upflow from the bifurcation point and the bifurcation point (solid line), and the bin downflow from the bifurcation point and the bifurcation point (dashed line), performed for head values of 1,10, and 100 (plus, pentagram and circle signs respectivly). As can be seen, while there is a distinct difference between the conductivity ratio before and after the bifurcation for both the value and trend, the head difference has little effect on them, pointing to the roubstness of the bifurcation charectrisitics.*

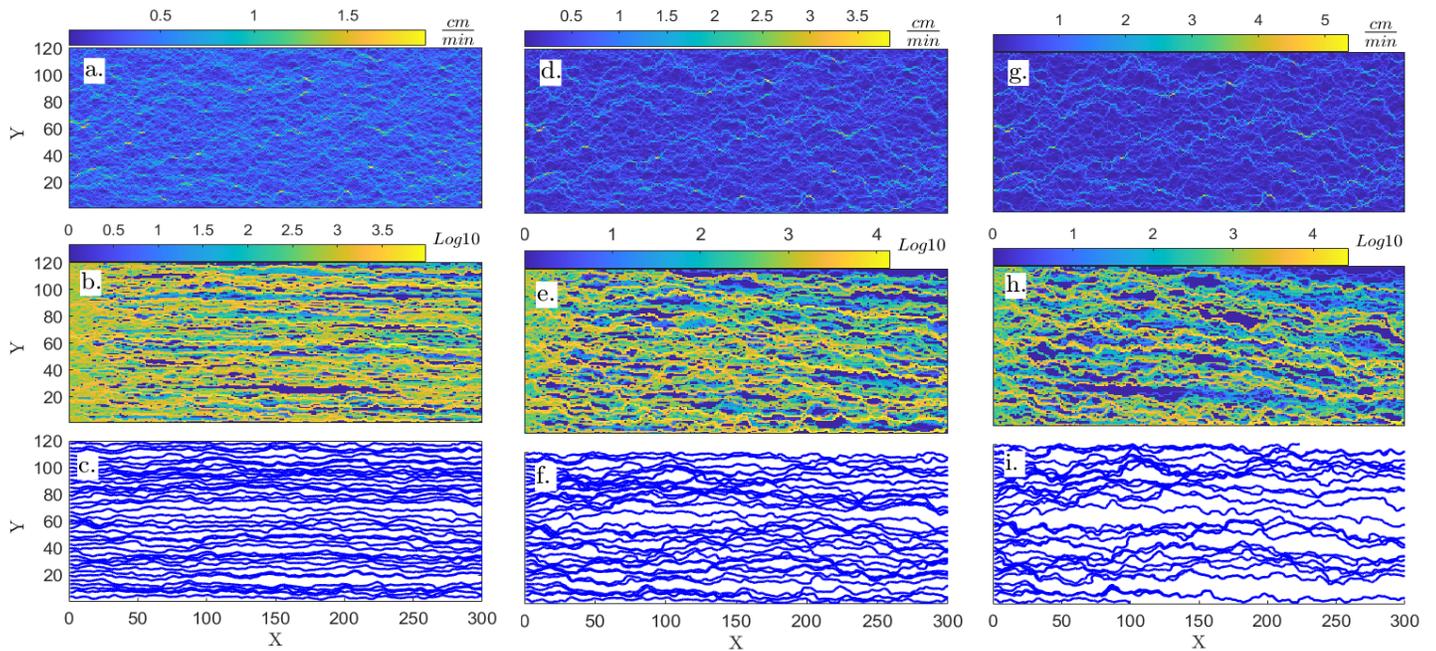

*Figure S4. The Velocity, particle visitation per bin, and streamline analysis of a single realization for variance 1 (a-c respectively), 3 (d-f respectively), and 5 (g-I respectively).*

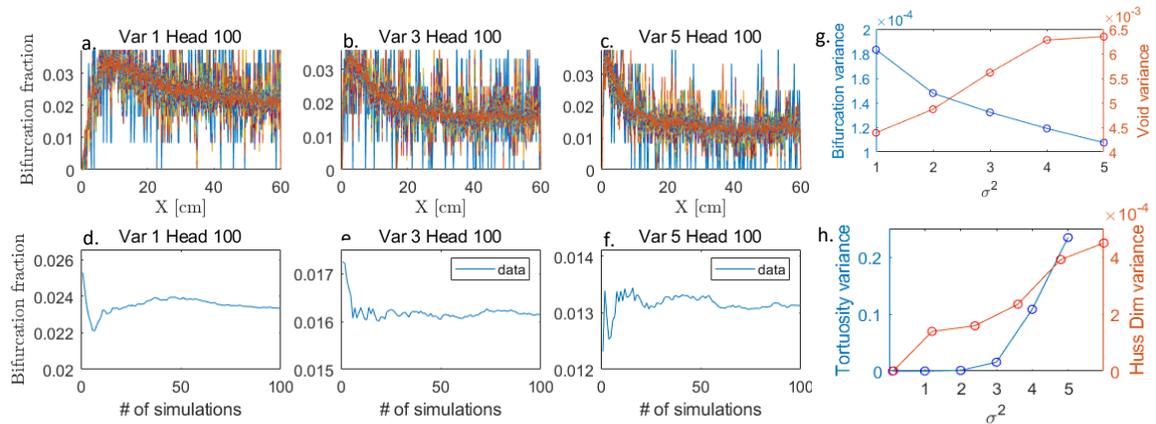

*Figure S5. a., b., and c. are an overlay of the bifurcation fraction, averaged after each additional 10 simulations, for variances 1, 3, and 5 respectively showing how rapid the convergence of the bifurcation fraction is, with every additional simulation. This is outlined for the asymptotical value of the bifurcation fraction, and averaged over every additional simulation (d., e., and f. for variances 1, 3, and 5 respectively). This robustness in the results is also apparent in the variance of the bifurcation and void fraction (figure g.), and in the tortuosity and Hausdorff dimension variances (figure h.).*

An additional analysis is shown for the fraction of distributed impervious values from 0.05 to 0.2 from the whole domain was performed. The nature of the impervious fraction does not allow flow through a fraction of the domain, leading to a linear increase in the void fraction, yet the same power-law distribution for bifurcation fraction, tortuosity, and Hausdorff dimension still emerges, as can be seen from our results in figure 1. Yet due to the nature of the binary system, namely that the ratio of impervious areas increases while there is but a single permeable value that is uncorrelated, the ratio of bifurcations increases as a power-law to an asymptote. From that asymptote, an increase in the impervious fraction will lead to the percolation threshold which was explored in many studies. While the parameters are not universal among different distributions, the scaling is, pointing to the results generality.

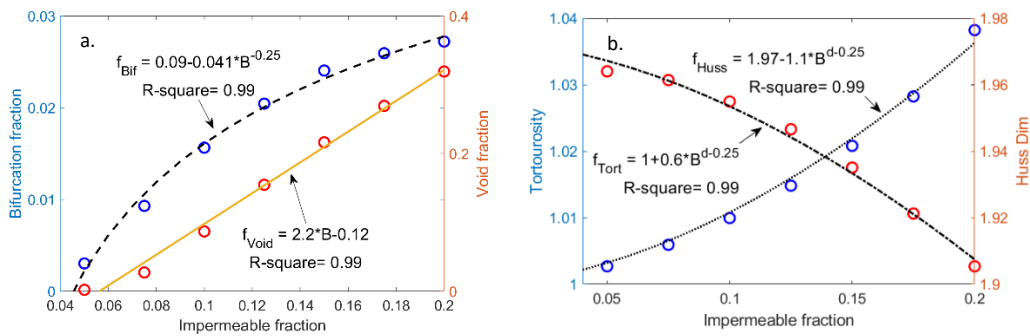

*Figure S6. a. Mean bifurcation fraction value, taken from the asymptotical value, and its power-law fit, as the impermeable area increase in blue circles and black dashed line respectively. Mean void fraction value as the impermeable area increases, and its power-law fit, in red circles and orange line respectively. b. Mean tortuosity value, calculated from particle trajectory, as the impermeable area increases, and its power-law fit, in blue circles and blue dotted line respectively. Mean Hausdorff fractal dimension as the impermeable area increases, and its power-law fit, in red circles and dotted- dashed line respectively.*

## References

Ababou, R., D. McLaughlin, L. W. Gelhar, and A. F. B. Tompson (1989), Numerical simulation of three-dimensional saturated flow in randomly heterogeneous porous media, Transport in Porous Media, 4(6), 549-565, doi: 10.1007/bf00223627.

Amooie, M. A., M. R. Soltanian, and J. Moortgat (2017), Hydrothermodynamic mixing of fluids across phases in porous media, Geophysical Research Letters, 44(8), 3624-3634, doi: doi:10.1002/2016GL072491.

Bear, J. (2013), Dynamics of fluids in porous media, Courier Corporation.

Berkowitz, B. (2021), HESS Opinions: Chemical transport modeling in subsurface hydrological systems–Space, time, and the holy grail of "upscaling", Hydrology and Earth System Sciences Discussions, 1-31.

Berkowitz, B., and H. Scher (1997), Anomalous transport in random fracture networks, Physical review letters, 79(20), 4038.

Bianchi, M., C. Zheng, C. Wilson, G. R. Tick, G. Liu, and S. M. Gorelick (2011), Spatial connectivity in a highly heterogeneous aquifer: From cores to preferential flow paths, Water Resources Research, 47(5).

Bolla Pittaluga, M., R. Repetto, and M. Tubino (2003), Channel bifurcation in braided rivers: Equilibrium configurations and stability, Water Resources Research, 39(3).

Bundt, M., F. Widmer, M. Pesaro, J. Zeyer, and P. Blaser (2001), Preferential flow paths: biological 'hot spots' in soils, Soil Biology and Biochemistry, 33(6), 729-738.
Carman, P. C. (1939), Permeability of saturated sands, soils and clays, The Journal of Agricultural Science, 29(2), 262-273.
Ciriello, V., V. Di Federico, M. Riva, F. Cadini, J. De Sanctis, E. Zio, and A. Guadagnini (2013), Polynomial chaos expansion for global sensitivity analysis applied to a model of radionuclide migration in a randomly heterogeneous aquifer, Stochastic environmental research and risk assessment, 27(4), 945-954.
Cirpka, O. A., and P. K. Kitanidis (2000), Characterization of mixing and dilution in heterogeneous aquifers by means of local temporal moments, Water Resources Research, 36(5), 1221-1236.
Cushman, J. H., and T. Ginn (1993), Nonlocal dispersion in media with continuously evolving scales of heterogeneity, Transport in Porous Media, 13(1), 123-138.
Domenico, P., and F. Schwartz 1990, Physical and chemical hydrogeology, edited, New York, John Wiley and Sons.
Dullien, F. A. (2012), Porous media: fluid transport and pore structure, Academic press.
Edery, Y. (2021), The Effect of varying correlation lengths on Anomalous Transport, Transport in Porous Media, 137(2), 345-364.
Edery, Y., H. Scher, and B. Berkowitz (2011), Dissolution and precipitation dynamics during dedolomitization, Water Resources Research, 47(8).
Edery, Y., S. Geiger, and B. Berkowitz (2016a), Structural controls on anomalous transport in fractured porous rock, Water Resources Research, 52(7), 5634-5643.
Edery, Y., D. Weitz, and S. Berg (2018), Surfactant variations in porous media localize capillary instabilities during Haines jumps, Physical review letters, 120(2), 028005.
Edery, Y., A. Guadagnini, H. Scher, and B. Berkowitz (2014), Origins of anomalous transport in heterogeneous media: Structural and dynamic controls, Water Resources Research, 50(2), 1490-1505.
Edery, Y., I. Dror, H. Scher, and B. Berkowitz (2015), Anomalous reactive transport in porous media: Experiments and modeling, Physical Review E, 91(5), 052130.
Edery, Y., M. Stolar, G. Porta, and A. Guadagnini (2021), Feedback mechanisms between precipitation and dissolution reactions across randomly heterogeneous conductivity fields, Hydrology and Earth System Sciences Discussions, 1-14.
Edery, Y., G. M. Porta, A. Guadagnini, H. Scher, and B. Berkowitz (2016b), Characterization of bimolecular reactive transport in heterogeneous porous media, Transport in Porous Media, 115(2), 291-310.
Eze, P. N., N. Madani, and A. C. Adoko (2019), Multivariate mapping of heavy metals spatial contamination in a Cu–Ni exploration field (Botswana) using turning bands co-simulation algorithm, Natural Resources Research, 28(1), 109-124.
Fasano, A., and F. Talamucci (2000), A comprehensive mathematical model for a multispecies flow through ground coffee, SIAM Journal on Mathematical Analysis, 31(2), 251-273.
Ferrari, A., J. Jimenez-Martinez, T. L. Borgne, Y. Méheust, and I. Lunati (2015), Challenges in modeling unstable two-phase flow experiments in porous micromodels, Water Resources Research, 51(3), 1381-1400.
Fiori, A., and I. Jankovic (2012), On preferential flow, channeling and connectivity in heterogeneous porous formations, Mathematical Geosciences, 44(2), 133-145.
Franssen, H. H., F. Stauffer, and W. Kinzelbach (2004), Influence of uncertainty of mean transmissivity, transmissivity variogram and boundary conditions on estimation of well capture zones, in geoENV IV—Geostatistics for Environmental Applications, edited, pp. 223-234, Springer.
Ghanbarian, B., A. G. Hunt, R. P. Ewing, and M. Sahimi (2013), Tortuosity in porous media: a critical review, Soil science society of America journal, 77(5), 1461-1477.
Gómez-Hernández, J. J., and A. G. Journel (1993), Joint sequential simulation of multigaussian fields, in Geostatistics Troia'92, edited, pp. 85-94, Springer.

Guadagnini, A., and S. P. Neuman (1999), Nonlocal and localized analyses of conditional mean steady state flow in bounded, randomly nonuniform domains: 1. Theory and computational approach, Water Resources Research, 35(10), 2999-3018, doi: 10.1029/1999wr900160.

Hagedorn, F., and M. Bundt (2002), The age of preferential flow paths, Geoderma, 108(1-2), 119-132.

Haggerty, R., S. A. McKenna, and L. C. Meigs (2000), On the late-time behavior of tracer test breakthrough curves, Water Resources Research, 36(12), 3467-3479.

Haggerty, R., S. W. Fleming, L. C. Meigs, and S. A. McKenna (2001), Tracer tests in a fractured dolomite: 2. Analysis of mass transfer in single-well injection-withdrawal tests, Water Resources Research, 37(5), 1129-1142.

Hencher, S. (2010), Preferential flow paths through soil and rock and their association with landslides, Hydrological processes, 24(12), 1610-1630.

Hunt, A. G., and M. Sahimi (2017), Flow, transport, and reaction in porous media: Percolation scaling, critical-path analysis, and effective medium approximation, Reviews of Geophysics, 55(4), 993-1078.

Huysmans, M., and A. Dassargues (2005), Review of the use of Péclet numbers to determine the relative importance of advection and diffusion in low permeability environments, Hydrogeology Journal, 13(5), 895-904.

Kanavas, Z., F. Pérez-Reche, F. Arns, and V. Morales (2021), Flow path resistance in heterogeneous porous media recast into a graph-theory problem, Transport in Porous Media, 1-16.

Kang, P. K., M. Dentz, T. Le Borgne, and R. Juanes (2011), Spatial Markov model of anomalous transport through random lattice networks, Physical review letters, 107(18), 180602.

Kesten, H. (1982), Proofs of Theorems 3.1 and 3.2, in Percolation Theory for Mathematicians, edited, pp. 168-197, Springer.

Koch, D. L., and J. F. Brady (1987), A non-local description of advection-diffusion with application to dispersion in porous media, Journal of Fluid Mechanics, 180, 387-403.

Kozeny, J. (1932), Die durchlässigkeit des bodens, Kulturtechniker, 35, 301-307.

Le Borgne, T., M. Dentz, and J. Carrera (2008), Lagrangian statistical model for transport in highly heterogeneous velocity fields, Physical review letters, 101(9), 090601.

Lenormand, R., C. Zarcone, and A. Sarr (1983), Mechanisms of the displacement of one fluid by another in a network of capillary ducts, Journal of Fluid Mechanics, 135, 337-353.

Levy, M., and B. Berkowitz (2003), Measurement and analysis of non-Fickian dispersion in heterogeneous porous media, J Contam Hydrol, 64(3-4), 203-226, doi: 10.1016/s0169-7722(02)00204-8.

Li, L., H. Zhou, and J. J. Gómez-Hernández (2011), Transport upscaling using multi-rate mass transfer in three-dimensional highly heterogeneous porous media, Advances in Water Resources, 34(4), 478-489.

Liao, K., and A. Scheidegger (1969), Branching-type models of flow through porous media, International Association of Scientific Hydrology. Bulletin, 14(4), 137-143.

Morales-Casique, E., S. P. Neuman, and A. Guadagnini (2006a), Non-local and localized analyses of non-reactive solute transport in bounded randomly heterogeneous porous media: Theoretical framework, Advances in water resources, 29(8), 1238-1255.

Morales-Casique, E., S. P. Neuman, and A. Guadagnini (2006b), Nonlocal and localized analyses of nonreactive solute transport in bounded randomly heterogeneous porous media: Computational analysis, Advances in water resources, 29(9), 1399-1418.

Morales, V. L., J.-Y. Parlange, and T. S. Steenhuis (2010), Are preferential flow paths perpetuated by microbial activity in the soil matrix? A review, Journal of Hydrology, 393(1-2), 29-36.

Moreno, L., and C. F. Tsang (1994), Flow channeling in strongly heterogeneous porous media: A numerical study, Water Resources Research, 30(5), 1421-1430.

Obi, I. S., K. M. Onuoha, O. T. Obilaja, and C. Dim (2020), Understanding reservoir heterogeneity using variography and data analysis: an example from coastal swamp deposits, Niger Delta Basin (Nigeria), Geologos, 26.

Pharoah, J., K. Karan, and W. Sun (2006), On effective transport coefficients in PEM fuel cell electrodes: Anisotropy of the porous transport layers, Journal of Power Sources, 161(1), 214-224.

Raveh-Rubin, S., Y. Edery, I. Dror, and B. Berkowitz (2015), Nickel migration and retention dynamics in natural soil columns, Water Resources Research, 51(9), 7702-7722.

Riva, M., M. De Simoni, and M. Willmann (2005), Impact of the choice of the variogram model on flow and travel time predictors in radial flows, in Geostatistics for Environmental Applications, edited, pp. 273-284, Springer.

Riva, M., L. Guadagnini, and A. Guadagnini (2010), Effects of uncertainty of lithofacies, conductivity and porosity distributions on stochastic interpretations of a field scale tracer test, Stochastic Environmental Research and Risk Assessment, 24(7), 955-970.

Riva, M., A. Guadagnini, D. Fernandez-Garcia, X. Sanchez-Vila, and T. Ptak (2008), Relative importance of geostatistical and transport models in describing heavily tailed breakthrough curves at the Lauswiesen site, Journal of contaminant hydrology, 101(1-4), 1-13.

Riva, M., A. Guadagnini, S. P. Neuman, E. B. Janetti, and B. Malama (2009), Inverse analysis of stochastic moment equations for transient flow in randomly heterogeneous media, Advances in Water Resources, 32(10), 1495-1507, doi: https://doi.org/10.1016/j.advwatres.2009.07.003.

Salamon, P., D. Fernàndez-Garcia, and J. J. Gómez-Hernández (2006), A review and numerical assessment of the random walk particle tracking method, Journal of contaminant hydrology, 87(3-4), 277-305.

Salamon, P., D. Fernàndez-Garcia, and J. Gómez-Hernández (2006), Modeling mass transfer processes using random walk particle tracking, Water Resources Research, 42(11).

Sánchez-Vila, X., and J. Carrera (2004), On the striking similarity between the moments of breakthrough curves for a heterogeneous medium and a homogeneous medium with a matrix diffusion term, Journal of Hydrology, 294(1-3), 164-175.

Sanchez-Vila, X., A. Guadagnini, and J. Carrera (2006), Representative hydraulic conductivities in saturated groundwater flow, Reviews of Geophysics, 44(3).

Shao, W., T. Bogaard, M. Bakker, and R. Greco (2015), Quantification of the influence of preferential flow on slope stability using a numerical modelling approach, Hydrology and Earth System Sciences, 19(5), 2197-2212.

Torelli, L., and A. E. Scheidegger (1972), Three-dimensional branching-type models of flow through porous media, Journal of Hydrology, 15(1), 23-35.

Tufenkji, N., and M. Elimelech (2004), Correlation equation for predicting single-collector efficiency in physicochemical filtration in saturated porous media, Environmental science & technology, 38(2), 529-536.

Webb, E. K., and M. P. Anderson (1996), Simulation of preferential flow in three-dimensional, heterogeneous conductivity fields with realistic internal architecture, Water Resources Research, 32(3), 533-545.

Willmann, M., J. Carrera, and X. Sánchez-Vila (2008), Transport upscaling in heterogeneous aquifers: What physical parameters control memory functions?, Water Resources Research, 44(12).

Xia, Y., J. Cai, W. Wei, X. Hu, X. Wang, and X. Ge (2018), A new method for calculating fractal dimensions of porous media based on pore size distribution, Fractals, 26(01), 1850006.

Yang, K., and K. Vafai (2011a), Transient aspects of heat flux bifurcation in porous media: an exact solution, Journal of heat transfer, 133(5).

Yang, K., and K. Vafai (2011b), Analysis of heat flux bifurcation inside porous media incorporating inertial and dispersion effects–an exact solution, International Journal of Heat and Mass Transfer, 54(25-26), 5286-5297.

Yeates, C., S. Youssef, and E. Lorenceau (2020), Accessing preferential foam flow paths in 2D micromodel using a graph-based 2-parameter model, Transport in Porous Media, 133(1), 23-48.

Yu, B., and P. Cheng (2002), A fractal permeability model for bi-dispersed porous media, International journal of heat and mass transfer, 45(14), 2983-2993.

Yu, B., and W. Liu (2004), Fractal analysis of permeabilities for porous media, AIChE journal, 50(1), 46-57.

Zehe, E., R. Loritz, Y. Edery, and B. Berkowitz (2021), Preferential Pathways for Fluid and Solutes in Heterogeneous Groundwater Systems: Self-Organization, Entropy, Work, Hydrology and Earth System Sciences Discussions, 1-28.

Zolezzi, G., W. Bertoldi, M. Tubino, G. Smith, J. Best, C. Bristow, and G. Petts (2006), Morphological analysis and prediction of river bifurcations, Braided rivers: process, deposits, ecology and management, 36, 233-256.